\documentstyle[osa,manuscript]{revtex}

\begin{document}
\title{Unusual magnetic relaxation behavior in La$_{0.5}$Ca$_{0.5}$MnO$_{3}$}
\author{J. L\'{o}pez*, P. N. Lisboa-Filho, W. A.\ C. Passos, W. A. Ortiz and F. M.
Araujo-Moreira}
\address{Grupo de Supercondutividade e Magnetismo,\\
Departamento de F\'{i}sica, Universidade Federal de S\~{a}o Carlos, Caixa\\
Postal - 676, S\~{a}o Carlos, SP, 13565-905, BRAZIL, \\
\*pjlopez@iris.ufscar.br}
\maketitle

\begin{abstract}
We present an extensive study of the time dependence of the magnetization in
a polycrystalline and low temperature charge ordered La$_{0.5}$Ca$_{0.5}$MnO$%
_{3}$ sample. After application and removal of a 5 T magnetic field, a
systematic variation of the magnetic relaxation rate from 10 K to 245 K was
found. At 195 K, the magnetization decreases in a very short time and after
that it increases slowly as a function of time. Moreover, between 200 and
245 K, an increase in magnetization, above the corresponding value just
after removing the 5 T magnetic field, was measured. This unusual behavior
was tested in several other relaxation procedures.

\end{abstract}

The low temperature charge ordered compound La$_{0.5}$Ca$_{0.5}$MnO$_{3}$\
has been intensively studied in the last years due to its very large
variations in resistivity, magnetization, and lattice parameters as a
function of temperature, magnetic field and isotope mass\cite{Radaelli}$^{,}$%
\cite{Zhao}$^{,}$\cite{Xiao}. Besides, this compound is particularly
interesting due to the coexistence and microscopic separation of two phases
at low temperatures: one ferromagnetic and one antiferromagnetic\cite
{Moritomo}. Recently, Huang et. al.\cite{Huang}, measured neutron powder
diffraction and magnetization in La$_{1-x}$Ca$_{x}$MnO$_{3}$ samples, with
x=0.47, 0.50 and 0.53. They argued the existence in all samples of\ a
paramagnetic-ferromagnetic transition, for a phase called F-I, at 265 K and
the formation of a second crystallographic phase, named A-II, below 230 K.
They also pointed out that phase A-II ordered antiferromagnetically with a
CE-type magnetic structure below 160 K\cite{Huang}. Moreover, Radaelli et.
al.\cite{Radaelli} measured a rapid change of the lattice parameters between
130 K and 225 K in La$_{0.5}$Ca$_{0.5}$MnO$_{3}$. This was associated with
the development of a Jahn-Teller distortion of the Mn-O octahedra, as well
as partial orbital ordering.

Relaxation experiments are a useful tool to study the dynamics of the charge
ordering phase due to the frustration in the spin equilibrium configuration.
Until now, attention has been focused in the relaxation of electrical
resistivity after a large change of an applied magnetic field, which induces
a transition from a ferromagnetic metallic state to a charge ordered
insulator phase or conversely\cite{Anane}$^{,}$\cite{Smolyaninova}. However,
systematic reports of magnetic relaxation curves (M(t)) in charge ordered
compounds are rare, possible because the abrupt jump seen in resistivity is
absent in magnetic measurements. Here, we present an extensive study M(t)
curves measured in a polycrystalline sample of La$_{0.5}$Ca$_{0.5}$MnO$_{3}$%
. We also performed similar measurements in a polycrystalline La$_{0.7}$Ca$%
_{0.3}$MnO$_{3}$ sample for comparison.

Polycrystalline samples of La$_{0.5}$Ca$_{0.5}$MnO$_{3}$ and La$_{0.7}$Ca$%
_{0.3}$MnO$_{3}$ were prepared by the standard procedures described
elsewhere \cite{Roy2}$^{,}$\cite{Paulo}. X-ray diffraction measurements
pointed out high quality samples. Magnetization measurements were done with
a standard MPMS-5S SQUID magnetometer. The relaxation measuring procedure
was the following: first, the sample was heated to 400 K in zero magnetic
field; second, the remanent magnetic field in the solenoid of the SQUID
magnetometer was seted to zero; third, the sample was cooled down to the
working temperature in zero magnetic field; fourth, an applied magnetic
field (H) was increased from 0 to 5 T at a rate of 0.83 T/minute and
remained applied for a waiting time t$_{w}$=50 s; fifth, H was decreased to
zero at the same rate; finally, when H was zero (we defined this moment as
t=0) the M(t) curve was recorded for more than 210 minutes. We have measured
the profile of the remanent magnetic field trapped in the superconducting
solenoid after increasing H to 5 T and the subsequent removal. Within the
experimental region the trapped magnetic field was lower than 1.1 mT.

Figure 1 shows the temperature dependence of the magnetization measured with
H=5 T for one of the La$_{0.5}$Ca$_{0.5}$MnO$_{3}$ samples studied. The
inset shows the same type of measurement with H=1.2 mT. In the first case
the measurement started at 2 K after a zero field cooling procedure, while
in the second (inset) started at 400 K. The large hysteresis, at magnetic
field as high as 5 T, is a clear evidence of the intrinsic frustration in
the equilibrium configuration of the spin system. The
paramagnetic-ferromagnetic transition around 225 K (minimum slope) is
independent of the applied field and the followed path. However, the
transition to the antiferromagnetic phase is path dependent and the Neel
temperature (T$_{N}$) decreases with the increase of H. Values of T$_{N}$,
calculated from the maximum slope in the curves, changed between 100 and 195
K. The shape of these curves is in agreement with previous reports in the
literature \cite{Xiao}$^{,}$\cite{Roy}.

Figure 2 shows magnetic relaxation measurements from 145 K to 195 K. To
facilitate the comparison between curves at different temperatures, the
magnetization in each case has been normalized to the corresponding value
just after removing the H=5 T. We will denote these curves as
m(t)=M(t)/M(0). As could be seen, the relaxation rate (mean slope of the
curves or $\Delta $m/$\Delta t$) increases systematically with increasing
temperatures from 150 K to 195 K. It is important to note that slopes here
are negatives. The ratio of magnetization change between t=$\infty $ (the
last measurement made) and t=0, m($\infty $), at 150 K is about 20 \%, while
at 195 K is only 0.9 \%. The increase in the relaxation rate between 150 and
195 K is in clear contrast with the behavior between 10 K and 150 K, where
the relaxation rate decreases\cite{J.López}. Included in figure 2 is the
curve corresponding to 145 K which illustrates this behavior. Measurements
of M versus H for this system \cite{Xiao}$^{,}$\cite{J.López} revealed the
complete destruction of the antiferromagnetic phase above 150 K with H=5 T.
This fact could be influencing the change of behavior of the relaxation rate
around 150 K.

The inset in figure 2 reproduces the relaxation curve before normalization
for 195 K. This curve is noticeable different from those at lower
temperatures. Here, the magnetization first decreases and, after
approximately 4 minutes, starts increasing with time. The inset in figure 3
also shows the M(t) curves at 200 K. In this case, no decrease in
magnetization was measured, but a monotonous increase with time
(approximately 62$\cdot $10$^{-5}\mu _{B}$ in 218 minutes) is observed. The
experiment at 200 K was repeated with\ a sample of the same compound having
only 9 \% of the original mass. A similar increase of magnetization with
time (88$\cdot $10$^{-5}\mu _{B}$ in 218 minutes) was found. These values
are about 10 times smaller than the magnetization measured with H=1.2 mT at
the same temperature (see inset in figure 1).

The main frame of figure 3 shows m(t) curves from 195 K to 245 K. Notice
that, differently from the previous figure, all curves, except the one at
195 K, show values above one. In other words, the magnetization increases
with time (curves have positive slopes) above the M(0) value in each case.
Furthermore, m($\infty $) values are systematically higher with higher
temperatures: from 0.9 \% at 195 K to 75 \% at 240 K. However, M(0)
decreases with higher temperatures, as shown in figure 1. The curve at 245 K
is also shown and presents a smaller m($\infty $) value with respect to the
one at 240 K, probably associated with the transition of the system to the
paramagnetic phase. Magnetic relaxation measurements were also done between
245 K and 350 K. In this temperature interval we did not find a systematic
variation of the relaxation rate. Nonetheless, above 260 K, the M(t) curves
always show the usual decreasing behavior with time.

As a small magnetic field is trapped (typically about 1 mT) in the
superconducting solenoid of the SQUID magnetometer, resulting after
successive applications and removals of H=5 T, we decided to check its role
in the unusual increase in magnetization. The temperature of 210 K was
chosen since the absolute value of magnetization is not too small and the
increase in normalized magnetization is higher than at 200 K. We repeated
the standard relaxation procedure, but instead of changing the applied
magnetic field from 5 T to H=0 mT (squares), we reduced it from 5 T to
H=-1.2 mT (circles) and H=-10 mT (triangles in the inset). In each case,
this last field remained applied during the whole relaxation measurement.
For H=-1.2 mT the actual field applied to the sample is almost zero. The
initial magnetization of the sample here is lower than in the H=0 mT case,
but it is still positive. However, the H=-10 mT is about 10 times higher
than the trapped flux in the SQUID solenoid and the actual applied field to
the sample is negative. The initial magnetization of the sample is lower
with H=-10 mT than in the previous cases and has a negative value (see
inset). Note that, contrary to what it is usually expected, the
magnetization increases with time even with applied negative magnetic
fields. The last measurements strongly support the point that is not the
small trapped magnetic field in the solenoid of the magnetometer which is
causing the unusual relaxation in\ La$_{0.5}$Ca$_{0.5}$MnO$_{3}$.

Next, we repeated the standard relaxation procedure changing t$_{w}$. Figure
5 shows three m(t) curves with t$_{w}$ equal to 50 s (squares), 500 s
(circles) and 5000 s (up triangles). Clearly, the normalized increase in
magnetization is higher the longer the 5 T magnetic field remains applied.
Values of M(0) also increase for longer t$_{w}$. A plausible explanation
could be that the remanent trapped field in the sample after removing the
H=5 T, which is higher for longer t$_{w}$, is causing the self-alignment of
the spins and the increase in magnetization. Therefore, these curves would
be reflecting information about the interactions in the La$_{0.5}$Ca$_{0.5}$%
MnO$_{3}$ sample.

Finally, we repeated the standard relaxation procedure for a La$_{0.7}$Ca$%
_{0.3}$MnO$_{3}$ polycrystalline sample. The ground state of this
composition\ is a single ferromagnetic phase with the
paramagnetic-ferromagnetic transition around 250 K. The magnetization, in
relaxation measurements both above and below the Curie temperature, always
decreased with time. These results for La$_{0.7}$Ca$_{0.3}$MnO$_{3}$ are in
agreement with a similar study reported for La$_{0.7}$Sr$_{0.3}$MnO$_{3}$%
\cite{Fisher}.

We conclude that is not only the ferromagnetic interaction that is causing
the increase in magnetization in La$_{0.5}$Ca$_{0.5}$MnO$_{3}$. In order to
understand this behavior we have to include other specific details of the
crystalline and magnetic structure of the charge ordered material. These
intriguing results could be a consequence of the competition between
ferromagnetic double exchange and antiferromagnetic superexchange coupling.
However, further studies in other charge ordering compounds, as well as
simultaneous measurements of magnetization, resistivity, etc. are desirable
to elucidate this unusual behavior.

We thank FAPESP, CAPES, CNPq and PRONEX for financial support. We also thank
professor P. Nozar for a helpful discussion.

\bigskip

Figure 1. Magnetization per Mn ion versus temperature measured with\ an
applied magnetic field of 5 T (1.2 mT in the inset) for a La$_{0.5}$Ca$%
_{0.5} $MnO$_{3}$ sample. Magnetization is given in Bohr magnetons per
manganese ion. Arrows and numbers show the direction of temperature sweep.

Figure 2. Normalized magnetic relaxation measurements after applying and
removing an applied magnetic field of 5 T. Time is shown in logarithmic
scale. The large arrow indicates the direction of increasing temperatures
and relaxation rates (slope less negative) between 150 K and 195 K. Also
shown is the curve for 145 K that presents a higher relaxation rate in
comparison with 150 K. The inset reproduces details of the curve at 195 K
with the corresponding error bars.

Figure 3. Normalized magnetic relaxation measurements after applying and
removing a magnetic field of 5 T. Time is shown in logarithmic scale. The
large arrow indicates the direction of increasing temperatures between 195 K
and 240 K. Also shown is the curve for 245 K that presents a smaller
increase of the normalized magnetization. The inset reproduces details of
the curve at 200 K with the corresponding error bars (same dimension of
circles), showing an unusual increase in the magnetization.

Figure 4. Magnetic relaxation curves at 210 K after repeating the standard
procedure but reducing the field from 5 T to H=0 mT (squares), H=-1.2 mT
(circles) and H=-10 mT (triangles in the inset). The initial magnetization
of the sample decreases with increasing negative magnetic field and all
curves show an increase in magnetization with time. Error bars are of the
same dimension of symbols for all curves.

Figure 5. Normalized magnetic relaxation measurements at 210 K after
applying a 5 T magnetic field, keeping it for a waiting time t$_{w}$= 50,
500 and 5000 s, and removing it. The increase in normalized magnetization is
higher the longer the magnetic field remains applied.


\begin{references}
\bibitem{Radaelli}  P. G. Radaelli, D. E. Cox, M. Marezio and S-W. Cheong,
Phys. Rev. B 55 (5) 3015 (1997)

\bibitem{Zhao}  Guo-meng Zhao, K. Ghosh, H. Keller and R. L. Greene, Phys.
Rev. B 59 (1) 81 (1999)

\bibitem{Xiao}  Gang Xiao, G. Q. Gong, C. L. Canedy, E. J. McNiff, Jr. and
A. Gupta, J. Appl. Phys. 81 (8) 5324 (1997)

\bibitem{Moritomo}  Y. Moritomo, Phys. Rev. B 60 (14) 10374 (1999)

\bibitem{Huang}  Q. Huang, J. W. Lynn, R. W. Erwin, A. Santoro, D. C.
Dender, V. N. Smolyaninova, K. Ghosh and R. L. Greene, Phys. Rev. B 61 (13)
8895 (2000)

\bibitem{Anane}  A. Anane, J. -P. Renard, L. Reversat, C. Dupas, P. Veillet,
M. Viret, L. Pinsard and A. Revcolevschi, Phys. Rev. B 59 (1) 77 (1999)

\bibitem{Smolyaninova}  V. N. Smolyaninova, C. R. Galley and R. L. Greene,
http://arxiv.org/abs/cond-mat/9907087, 6 Jul 1999

\bibitem{Roy2}  M. Roy, J. F. Mitchell, A. P. Ramirez and P. Schiffer, Phys.
Rev. B 58 (9) 5185 (1998)

\bibitem{Paulo}  Lisboa Filho, P. N., S. M. Zanetti, E. R. Leite and W.
A.Ortiz, Materials Letters 38 (4), 289 (1999)

\bibitem{Roy}  M. Roy, J. F. Mitchell and P. Schiffer,
http://arxiv.org/abs/cond-mat/0001064, 6 Jan 2000.

\bibitem{J.López}  J. L\'{o}pez, P. N. Lisboa-Filho, W. A.\ C. Passos, W. A.
Ortiz and F. M. Araujo-Moreira, http://arXiv.org/abs/cond-mat/0004460, 26
Apr 2000 (2000)

\bibitem{Fisher}  L. M. Fisher, A. V. Kalinov, S. E. Savel 'ev, I. F.
Voloshin and A. M. Balbashov, J. Phys.:Condens. Matter 10, 9769 (1998)
\end{references}
\end{document}